\documentclass[twocolumn,aps]{revtex4}
\usepackage[dvipdf]{graphicx}
\begin{document}
\newcommand{\avg}[1]{\langle{#1}\rangle}
\newcommand{\Avg}[1]{\left\langle{#1}\right\rangle}
\def\be{\begin{equation}}
\def\ee{\end{equation}}
\def\bc{\begin{center}} 
\def\ec{\end{center}}
\def\bea{\begin{eqnarray}}
\def\eea{\end{eqnarray}}
\def\bwt{\begin{widetext}}
\def\ewt{\end{widetext}}
\def\ra{\rightarrow}
\def\ba{\backslash}
\def\pa{\partial}
\title{ Flux distribution of metabolic networks close to  optimal
  biomass production}
\author{Ginestra Bianconi}

\affiliation{ The Abdus Salam International Center for Theoretical Physics, 
Strada Costiera 11,34014 Trieste,Italy}
\begin{abstract}
We study a statistical model describing the steady state distribution 
of the fluxes in a metabolic network. The resulting 
model on continuous variables can be solved by the cavity method.
In particular analytical tractability is possible solving the cavity
equation over an ensemble of networks with the same degree distribution
of the real metabolic network. The flux distribution that optimizes production of 
biomass has a fat tail with a power-law exponent independent on the structural 
properties of the underling network. These results are in complete agreement with 
the  Flux-Balance-Analysis outcome of the same system and in qualitative
agreement with the experimental results.
\end{abstract}

\maketitle

Recently large attention has been addressed by the physics community
to critical phenomena \cite{Doro_rev} in complex networks 
\cite{Review1,Review2,Review3,Review4,Review5,Review6}.
The complex topologies, usually characterized
by non-Poisson degree distributions, have large effect on the
critical point and critical exponents of the dynamical models defined
on them. The Ising model \cite{Ising1,Ising2,Ising3},
the epidemic spreading \cite{Vesp} and the synchronization
dynamics \cite{motter} are examples of dynamics models, where the complex 
structure has strong implications.
Furthermore in the last decade we have assisted to a big breakthrough in system biology, the 
interdisciplinary field that studies the biological problems going beyond the 
single biomolecule framework, with the
description of the intertwined reactions between the constituents of
the cell in terms of networks. This has generated a new 
theoretical framework in which new biological statistical findings have 
been formulated \cite{metabolic,Sharan,net_bio}.
In system biology there was also the fast development of ``in silico''
biology in which new experiments are simulated and the predictions are made
to stimulate further experimental confirmations of the phenomena.
A key example of a biological system in which the network picture is crucial and
the ``in silico'' biology has made relevant advances, is the prediction of
the growth rate of single cells of different organisms and the study of
the metabolic networks.
Two key advances in this field have been the full characterization of 
the chemical reactions \cite{fba} 
for a series of model organisms, as different
strain of {\it Escherichia coli} and {\it Saccaromyces cerevisiae}
(see for example the BIGG database \cite{BIGG}), and the application of the techniques of 
linear programming for the study of the flux of the reaction, an extension which
goes under the name of Flux-Balance-Analysis \cite{fba}.

The set of stoichiometric interactions in the cell can be represented as a
network whose nodes are of two types, the metabolic substrates of
metabolites and the nodes representing the reactions. This bipartite
network goes under the name of factor graph. In a factor graph, to
each reaction $i$ is assigned a flux variable $s_i$ and to each metabolite $\mu$ 
is assigned a steady state
condition for the production/consumption of the metabolites.
The structure of the metabolic network has a projection on the
metabolites which has a power-law degree distribution \cite{metabolic} and a 
hierarchical structure \cite{modular,PRL_jap}.
In the metabolic networks, to each reaction corresponds an enzyme which
regulates the rate of each reaction and modulates
the flux of the reactions. Consequently the maximal flux rate is fixed
by the maximal enzyme concentration inside the cell.
 Solving the non-linear mass-law equations is a hard
problem in networks of thousand of nodes.
To overcome this problem in Flux-Balance-Analysis for each reaction
a new variable, its flux, is introduced. Each flux 
includes all the dynamical effects 
associated to each reaction of the organism. The Flux-Balance-Analysis 
\cite{fba,Palsson} considers the steady state of
the dynamics which optimizes the production of the biomass by linear
programming.

The underline assumption of Flux-Balance-Analysis, that the cell
organisms optimize the biomass is very well confirmed by experimental
results \cite{Palsson} conducted in rich medium but is not fully supported by
experiments measuring the growth rate of single knockout strains.
For this reason other algorithms have been designed relaxing this condition. 
 \cite{MOMA,ROOM}.

In this paper we will study the flux distribution in metabolic networks that has a 
heavy tail as found in experiments \cite{Experiments} and in Flux-Balance-Analysis 
\cite{Almaas}
predictions in ${\it Echerichia\ coli}$.
In particular we add to the description of the metabolic
networks some theoretical statistical mechanics insights using the
cavity method \cite{Yedida,Doro_rev}.  Different theoretical models have been 
already proposed \cite{Monasson, Mulet,NHM} for the flux distribution but neither 
of them has been able to theoretically predict the
outcome of the experiments or of the Flux-Balance-Analysis
calculations. 
Here we will relate the power-law exponent of the flux
distribution with the steady distribution as an indication of
criticality. In fact it can be predicted by assuming that the network
is optimizing the biomass production. Thus we will
characterize this optimized state doing an asymptotic expansion of the
cavity equation close to the critical point and we
will measure the critical exponents corresponding to this critical transition. 
The method which we formulate here is a new method to solve the cavity
equation on continuous variables defined on a compact interval of the
real axis and it can be extended to other critical phenomena on complex
networks and continuous variables defined in a limited interval.
We find that 
the distribution of the fluxes present in the optimized state develops
a power-law tail with
exponents that are independent on the structure of the underlying
network, a new scenario in which the complex topology does not affect the
critical state of the cell.
The power-law exponent that we find is in full agreement with the
results of Flux-Balance-Analysis\cite{Almaas} and only partially in agreement with
the experimental results \cite{Experiments}

{\it The model-} 
The metabolic network has a bow tie structure \cite{PRL_jap} : the metabolites are
divided into input metabolites which are provided by the environment,
the output metabolites which provide the biomass and intermediate
metabolites.
The stoichiometric matrix is given by $((\xi^{\mu}_i))$ where
$\mu=1,\dots,M$ indicates the number of metabolites and $i=1,\dots,N$
the number of reactions and the sign of $\xi^{\mu}_i$ indicates if the
metabolite $\mu$ is an input or output metabolite of the reaction $i$.
In the Flux-Balance-Analysis method we assume that each intermediate
metabolite has a concentration $c^{\mu}$ at steady state, i.e.
\be
\frac{d c_{\mu}}{dt}=\sum_j \xi_{j,\mu} s_i=0
\label{ss.eq}
\ee
where $s_i$ is the flux of the metabolic reaction $i$.
For the metabolites present in the environment and the metabolites
giving rise to the biomass production we can fix the 
incoming flux given by
\be
\frac{d c_{\mu}}{dt}=\sum_j \xi_{j,\mu} s_i=g_{\mu}^{in/out}.
\label{ss2.eq}
\ee
We have already mentioned that the fluxes have some biological limitations.
To describe these limitations we assume that the fluxes $s_i\in[0,L]$.
 
The volume of solutions $V$ of this problem is given by 
\be
V= \int_0^L\ldots \int _0^L \prod_{i=1}^{N} ds_i
\prod_{\mu}\delta(\sum_j \xi_{j,\mu} s_i-g_{\mu}).
\label{z.eq}
\ee

 \begin{figure}
\centering{
 \includegraphics[width=7.5cm, height=5.5cm]{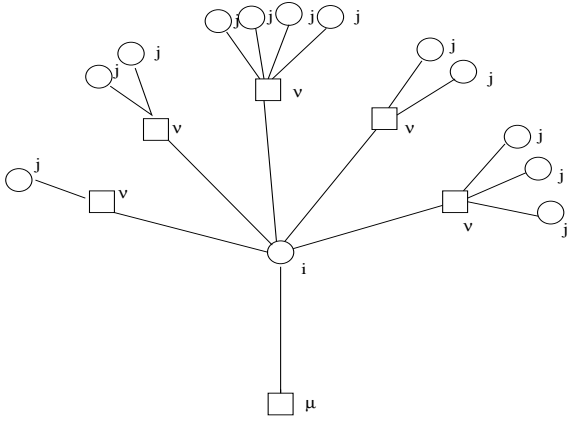}
 \caption{The cavity tree ${\cal C}_{\mu}$. If the metabolite $\mu$ is removed
 from the metabolic factor graph, to each reaction $i$ in which it
 takes part we can assign a flux $s_i$ with the cavity distribution $p_{i\ra \mu}(s_i)$.}
 \label{tree}
}
\end{figure}

{\it Belief Propagation-}
Let us assume that the factor graph of the metabolic network has a local
tree like structure as shown in figure \ref{tree}. In this assumption
Belief Propagation (BP) equations are able to fix the probability
distribution of the metabolic fluxes with the measure defined in 
$(\ref{z.eq})$.
BP equations are defined on cavity graphs. The cavity graphs ${\cal
 C}_{\mu}$ is the full factor graph of the metabolic network in the absence of 
metabolite $\mu$.
In ${\cal C}_{\mu}$ the flux $s_i$ of a reaction $i$ in which $\mu$
is reacting has a {\it cavity distribution} $p_{i\ra\mu}(s_i)$.
Expressing $p_{i\ra\mu}(s_i)$ in terms of the cavity distribution
$p_{j\ra\nu}(s_i)$ where $i,j,\mu,\nu$ are defined in figure
$\ref{tree}$, we get the BP equations:
\bea
p_{i\ra \mu}(s_i)&=& \frac{1}{C_{i,\mu}}\prod_{\nu\in N(i)\ba \mu}\prod_{j\in 
N(\nu)\ba
i} \left[\int ds_j p_{j\ra\nu}(s_j)\right] \nonumber \\
& &\prod_{\nu\in N(i)\ba \mu}
\delta(\sum_j \xi_{j,\nu}s_j +\xi_{i,\nu}s_i -g_{\nu}).
\label{BP}
\eea

\begin{figure}
\includegraphics[width=80mm, height=60mm]{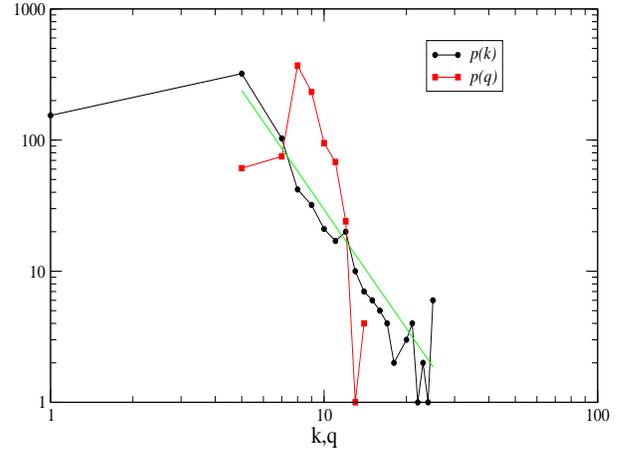}
\caption{\label{denk.fig} The degree distribution of $p(k)$ and $p(q)$
  for {\it Escherichia coli}, data taken form the BIGG database
  \cite{BIGG}.The line indicates the power-law $p(k)=k^{-\gamma}$ with $\gamma=3.0$. } 
\end{figure}

Solving the BP equations for the cavity distributions the marginal
probability of a flux $s_i$ is given by
\bea 
p_{i}(s_i)&=& \frac{1}{C_i}\prod_{\nu\in N(i)}\prod_{j\in N(\nu)\ba
i} \left[\int ds_j p_{j\ra\nu}(s_j)\right]\nonumber \\
&& \prod_{\nu\in N(i)}
\delta(\sum_j \xi_{j,\nu}s_j +\xi_{i,\nu}s_i -g_{\nu}).\nonumber
\eea
The distribution of the fluxes producing/consuming the metabolite
$\mu$ , i.e.
$\underline{S}_{\mu}=\{s_i\}_{i\in N(\mu)}$ is given by
\be
p_{\mu}(\underline{S}_{\mu})=\frac{1}{C_{\mu}}\delta(\sum_j \xi_{j,\mu}s_j-
g_{\mu})\prod_{j\in N(\mu)}p_{j\ra \mu}(s_j)
\ee 
The entropy of the metabolic network can be expressed as 
\bea
\Sigma&=&-\sum_{\mu}\int \prod_{i\in N(\mu)}ds_i
p_{\mu}(\underline{S}_{\mu})\log p_{\mu}(\underline{S}_{\mu})+\nonumber \\
& &\sum_i (k_i-1)\int ds_i
p_i({s}_{i})\log p_i({s}_{i}).
\eea

{\it The optimized state of a random metabolic network-}
We assume that the metabolic network is a random graph with $M$
metabolites with degree distribution $p(k)$ and $N$ reaction nodes
with degree distribution $p(q)$. In this network the total number of
links is given by $N \avg{q}=M\avg{k}$.
In Figure $\ref{denk.fig}$ we show an example of $p(k)$ and $p(q)$
distributions for {\it Escherichia coli}.
The $p(k)$ degree distribution for this organism is power-law $p(k)\sim
k^{-\gamma}$ with an exponent $\gamma\simeq 3$ while the $p(q)$
distribution is much more picked.
In different organisms the distribution of $p(q)$ and $p(k)$ do change but
the general scenario of power-law $p(k)$ distribution and finite scale
$p(q)$ distribution remains unchanged.

To solve the BP equations $(\ref{BP})$ in this random ensemble of
graphs
with given $p(k)$ and $p(q)$ degrees distributions  
we introduce the quantity
\be
P(s)=\frac{1}{\avg{q}N}\sum_{i,\mu\in\pa 
i}\Avg{\overline{p_{i\ra \mu}(s)}}_{\rho(g)}
\ee
where the overline $\overline{\ldots}$ indicates the average over random sign of 
the
reactions and the outgoing and ingoing flux distribution  i.e. over the 
distribution for the $\{\xi\}$
$P(\{\xi\})=\prod_i\prod_{\nu\in\pa
 i}\frac{1}{2}\left[\delta(\xi_{i,\nu}-1)+\delta(\xi_{i,\nu}+1)\right]$ and
 $\avg{\ldots}$ indicates the average over the probability distribution $\rho(g)$ of the $g$'s.
In particular we take 
\be
\rho(g)=p_1\delta(g+g_1)+p_2\delta(g-g_2)+(1-p_1-p_2)\delta(g)
\ee
where $p_1$ and $p_2$ indicate respectively the fraction of input
metabolites and the fraction of metabolites in the biomass definition and where 
$g_1$ is fixed by the environment
conditions and $g_2$ is the rate of the biomass.

The Fourier transform of $P(s)$ is the function $\chi(w)$, 
\bea
\chi(w)=\int ds e^{iws}P(s). 
\eea
where $w=\frac{2\pi}{L}n$. The function $\chi(w)$  satisfies the self consisted equation
\begin{widetext}
\bea
\chi(w)&=&\sum_{q}\frac{qp(q)}{\avg{q}}
\prod_{\nu=1}^{q-1}\left\{\sum_{k_{\nu}}\frac{k_{\nu}p(k_{\nu})}{\avg{k}}\sum_{\{n_{\nu}\}}
\sum_{\omega_{\nu}}\left[\frac{1}{2}(\chi(-\omega_{\nu})+\chi(\omega_{\nu}))\right]^{k_{\nu}-1}\right\} \Avg{\frac{e^{-
i\sum_{\nu}\omega_{\nu}{g_{\nu}}}}{C(\{k_{\nu}\})}}_{\rho(g)}\delta(\sum_{\nu}\omega_{\nu}(-1)^{n_{\nu}}-w).\nonumber
\label{sf}
\eea
\end{widetext}
where $\avg{\cdot}$ indicates the average over the distribution of the
incoming/outgoing fluxes $g$ and where and we have assumed $L\gg1$
allowing for a continuous variables $\omega_{\nu}$, $w$.
\bea
C(\{k_{\nu}\})&\simeq&\prod_{\nu=1}^{q-1} \left\{\sum_{
\omega_{\nu}}\left[\frac{1}{2}(\chi(-
\omega_{\nu})+\chi(\omega_{\nu}))\right]^{k_{\nu}-1}\right\}\nonumber
\\
&& e^{-i\sum_{\nu}{\omega_{\nu}}g_{\nu}}\delta(\sum_{\nu}\omega_{\nu}(-1)^{n_{\nu}}).
\eea
 The equation for $\chi(w)$ will have solutions until the rate of
 biomass production ${g}_2<G_c$ with
$G_c$ been corresponding to the maximal allowed biomass production of
the metabolic network.

Close to the critical point ${g}_2\simeq G_c$ we suppose that the
distribution $P(s)$ will scales as 
\be
P(s)=s^{-\tau}\Phi(s|{g}_2-G_c|^{\sigma})
\ee
where $\Phi$ is a scaling function.
In the limit of large $L$ we can assume that $w$ (together with the $\omega_{\nu}$) is a continuous
variable and we can  do an asymptotic expansion is reflected the scaling behavior of 
$\chi(w)$, i.e. 
\bea
\chi(w)=1-|w|^{\tau-1}h(w/|{g}_2-G_c|^{\sigma}).
\eea 
Solving then the self consistent equation for $\chi(w)$
Eq. $(\ref{sf})$ for analytic distribution $p(q)$
and the distribution of the metabolites connectivity decaying like a
power-law $p(k)\sim k^{-\gamma}$.
Close to the phase transition we have
\begin{widetext}
\bea
C(\{k_{\nu}\})&\simeq&\int\dots \int
\prod_{\nu=1}^{q-1}d\omega_{\nu}\left[1-(k_{\nu}-1)|\omega_{\nu}|^{\tau-
1}\mbox{Re}\ h
+\frac{1}{2}(k_{\nu}-1)(k_{\nu}-2)|\omega_{\nu}|^{2(\tau-1)}(\mbox{Re}\
h )^2\right] \nonumber \\ && e^{-i\sum_{\nu}{\omega_{\nu}}g_{\nu}}\delta(\sum_{\nu}\omega_{\nu}(-
1)^{n_{\nu}})\nonumber
\\ &\simeq&1+\sum_{\nu,\nu'}\left[A_1(g_{\nu},g_{\nu'})
 (k_{\nu}-1)(k_{\nu'}-1)+A_2(g_{\nu},g_{\nu'})(k_{\nu}-1)(k_{\nu}-2)(k_{\nu'}-
1)\right]\nonumber \\
&&-\sum_{\nu, nu',\nu''}A_3(g_{\nu},g_{\nu'},g_{\nu''})(k_{\nu}-1)(k_{\nu'}-1)(k_{\nu''}-1)
\eea

where $A_1$, $A_2$ and $A_3$ are linear functions of $g_{\nu}$, $g_{\nu'}$ and $g_{\nu''}$. If we
develop (\ref{sf}) around the point $w=0$, we get

\bea
\chi(w)&=&\sum_{q}\frac{qp(q)}{\avg{q}}
\prod_{\nu=1}^{q-1}
\left\{ \sum_{k_{\nu}}\frac{k_{\nu}p(k_{\nu})}{\avg{k}}\sum_{\{n_{\nu}\}}\int
d\omega_{\nu}\left[1-(k_{\nu}-1)|\omega_{\nu}|^{\tau-1}(\mbox{Re}\ h)+\frac{1}{2}k_{\nu}(k_{\nu}-
1)|\omega_{\nu}|^{2(\tau-1)}(\mbox{Re}\ h)^2\right]\right\}
\nonumber \\
& & \times \Avg{\frac{(1-i\sum_{\nu}\omega_{\nu}{g_{\nu}})}{C(\{k_{\nu}\})}}_{\rho(g)}\delta(\sum_{\nu}\omega_{\nu}(-1)^{n_{\nu}}-w).
\label{chi2}
\eea
\end{widetext} 
Since the sum over the connectivity $k_{\nu}$ are convergent in a
sparse network, all metabolic networks have for the flux distribution the mean
field exponents $\tau=3/2$ and $\sigma=2$ as long as the hypothesis of
Flux-Balance-Analysis are satisfied.

The entropy goes like 
\be 
\Sigma=|g_2-G_c|^{\alpha}
\ee
with $\alpha=\sigma(\tau-1)=1$

Therefore in the physical range for each degree distribution $p(q)$ or 
$p(k_{\nu})$ \cite{metabolic} the
predicted power-law critical exponent for the flux distribution is
$\tau=3/2$ in good agreement with the 
Flux-Balance calculations \cite{Almaas}

{\it Conclusion-}
We have presented a statistical mechanics approach to study the steady state
distribution of the fluxes in the metabolic network assuming
optimization of the biomass.
The analytic treatment
finds a distribution of the fluxes which is a power-law with an
mean-field exponent $\tau=3/2$ independent on the structure of the
metabolic network.
The method can be generalized to other critical phenomena 
defined on continuous variables on a finite interval, and work in this direction 
is in progress.
\\
We acknowledge interesting discussions with Riccardo Zecchina, 
this work was supported by IST STREP GENNETEC contract No.034952.

\end{document}